\documentclass[12pt]{article}
\usepackage{type1cm,amsmath}
\usepackage[psamsfonts]{amssymb}
\usepackage{cite}
\numberwithin{equation}{section}

\newcommand{\del}{\partial}
\newcommand{\ib}{\bar{\imath}}
\newcommand{\jb}{\bar{\jmath}}
\newcommand{\kb}{\bar{k}}
\newcommand{\lb}{\bar{\ell}}
\newcommand{\mb}{\bar{m}}
\newcommand{\nb}{\bar{n}}
\newcommand{\wb}{\bar{w}}
\newcommand{\zb}{\bar{z}}
\newcommand{\Omegab}{\bar{\Omega}}
\newcommand{\vp}{\varphi}
\newcommand{\ve}{\varepsilon}
\newcommand{\slsh}[1]{/ \kern -0.5em #1}
\newcommand{\vac}{|0\rangle}
\newcommand{\vact}{|\tilde 0\rangle}

\DeclareMathOperator*{\im}{{\rm Im}}
\newcommand{\nn}{\nonumber\\}
\newcommand{\Ncal}{{\cal N}}

\newcommand{\ta}{\theta}
\newcommand{\tb}{\bar{\theta}}
\newcommand{\deldel}[2]{\frac{\del #1}{\del #2}}
\newcommand{\dlr}{\overset{\leftrightarrow}{\del}}
\newcommand{\Qb}{\bar{Q}}
\newcommand{\Db}{\bar{D}}
\newcommand{\psb}{\bar{\psi}}
\newcommand{\etb}{\bar{\eta}}
\newcommand{\Phb}{\bar{\Phi}}
\newcommand{\Fb}{\bar{F}}
\newcommand{\Wb}{\bar{W}}

\newcommand{\pha}{z}
\newcommand{\phb}{\bar{z}}

\begin{document}

\thispagestyle{empty}

\begin{flushright}
 \begin{tabular}{l}
 UT-02-55\\
 {\tt hep-th/0210262}\\
 \end{tabular}
\end{flushright}

 \vfill
 \begin{center}
 \font\titlerm=cmr10 scaled\magstep4
 \font\titlei=cmmi10 scaled\magstep4
 \font\titleis=cmmi7 scaled\magstep4
 \centerline{\titlerm D-branes in PP-Waves and} 
 \vskip 3mm
 \centerline{\titlerm Massive Theories on Worldsheet with Boundary}

 \vskip 2.0 truecm
\noindent{ \large Yasuaki Hikida and Satoshi Yamaguchi} \\
{\sf hikida@hep-th.phys.s.u-tokyo.ac.jp~,~
yamaguch@hep-th.phys.s.u-tokyo.ac.jp}
\bigskip

 \vskip .6 truecm
 {
 {\it Department of Physics,  Faculty of Science, University of Tokyo \\
  Hongo 7-3-1, Bunkyo-ku, Tokyo 113-0033, Japan} 
 }
 \vskip .4 truecm

 \end{center}

 \vfill
\vskip 0.5 truecm

\begin{abstract}

We investigate the supersymmetric D-brane configurations in
the pp-wave backgrounds proposed by Maldacena and Maoz.
We study the surviving supersymmetry in a D-brane configuration
from the worldvolume point of view.
When we restrict ourselves to the background with $\Ncal=(2,2)$ 
supersymmetry and no holomorphic Killing vector term, 
there are two types of supersymmetric D-branes: A-type and B-type.
An A-type brane is wrapped on a special Lagrangian submanifold, and the
imaginary part of the superpotential should be constant on its worldvolume.
On the other hand, a B-type brane is wrapped on a complex submanifold, and
the superpotential should be constant on its worldvolume. 
The results are almost consistent with the worldsheet theory
in the lightcone gauge.
The inclusion of gauge fields is also discussed and found BPS
D-branes with the gauge field excitations.
Furthermore, we consider the backgrounds with holomorphic Killing
vector terms and $\Ncal=(1,1)$ supersymmetric backgrounds.
 
\end{abstract}
\vfill
\vskip 0.5 truecm

\newpage
\setcounter{page}{1}


\section{Introduction}

The string theory on a Ramond-Ramond background is an interesting problem
since the RNS formalism cannot be applied to the RR background and
we should use the Green-Schwarz (GS) formalism. 
This class of background has received attention
also because the most typical background for the AdS/CFT 
correspondence --- $AdS_5 \times S^5$ --- includes the RR fields.

The simplest example of RR-background is the maximally supersymmetric
IIB plane wave with RR 5-form flux \cite{Blau:2001ne,Blau:2002dy,Blau:2002mw}. 
The string theory on this background becomes massive free theory
in the lightcone gauge, and can be solved \cite{Metsaev:2001bj,Metsaev:2002re}.
The string theory on this background is claimed to correspond to the
subsector of
the 4-dimensional $\Ncal=4$
Yang-Mills theory \cite{Berenstein:2002jq}.
Recently, a larger class of supersymmetric pp-wave backgrounds are
investigated \cite{Maldacena:2002fy} (see also
 \cite{Berkovits:2002vn,Russo:2002qj,Bonelli:2002fs,Berkovits:2002rd}).
The string theories
on these backgrounds are proposed to become supersymmetric Landau-Ginzburg
theories in the lightcone gauge.

In this paper, we consider D-branes in these RR backgrounds. 
We expect that we can understand the GS open strings by studying the
D-branes. The D-branes also play an important role in the holography; 
They are supposed to correspond to the defects in CFT
\cite{Karch:2000gx,DeWolfe:2001pq,Bachas:2001vj,Erdmenger:2002ex,Mateos:2002bu,Yamaguchi:2002pa}. 
The recent works on the D-branes in pp-wave backgrounds are
\cite{Billo:2002ff,Dabholkar:2002zc,Berenstein:2002zw,%
   Chu:2002in,Lee:2002cu,Kumar:2002ps,%
   Skenderis:2002vf,Balasubramanian:2002sa,Takayanagi:2002je,%
   Bain:2002nq,Alishahiha:2002rw,Bergman:2002hv,Seki:2002wz,Mateos:2002an,%
   Pal:2002vx,Naculich:2002fh,Michishita:2002jp,Bain:2002tq,Alishahiha:2002nf,%
   Biswas:2002yz,Takayanagi:2002nv,Nayak:2002ty}.

In the RNS strings in the conformal gauge, consistent D-branes
are expressed as the boundary conditions
preserving the superconformal symmetry. 
Then, what is the condition of consistent D-branes 
on the GS strings in the lightcone gauge?
The boundary conditions preserving the supersymmetry in Landau-Ginzburg theory
are known\cite{Hori:2000ck}. We try to see whether these boundary
conditions correspond to consistent D-branes.
We consider the supersymmetric D-brane configurations from the viewpoint
of the worldvolume theory, 
then we compare them to the worldsheet theories with supersymmetric
boundary conditions.

We first consider the ${\cal N}=(2,2)$ supersymmetric case\footnote{
In this paper, we express the type of supersymmetry in terms of two
dimensional worldsheet theory.
} without
holomorphic Killing vector term. 
The kappa symmetry projection on worldvolume theory is used to examine
the supersymmetry and the following results are obtained.
The supersymmetric D-branes are classified into the two types according
to the preserved supersymmetry: A-type and B-type.
An A-type D-brane should be wrapped on a special Lagrangian submanifold and
the imaginary part of the superpotential should be constant on the
D-brane worldvolume. 
On the other hand, a B-type D-brane should be wrapped on a complex
submanifold and the superpotential should be constant on the D-brane
worldvolume. 
These results can be reproduced by the 
${\cal N} = (2,2)$ supersymmetric Landau-Ginzburg theory with the 
boundary conditions preserving worldsheet supersymmetry.
One exception is that the A-branes are wrapped on Lagrangian
(not necessarily {\em special} Lagrangian) submanifolds in this case.

Then we consider more general D-brane configurations.
Since the D-brane equation of motion may suggest the inclusion of the gauge
fields, we consider the D-branes with the gauge field excitation. 
The D-branes in the background with holomorphic Killing vector term and
${\cal N} = (1,1)$ supersymmetric background are also investigated.

The organisation of this paper is as follows: In section \ref{section-review},
we review the supersymmetric pp-wave backgrounds constructed in
\cite{Maldacena:2002fy}. We explain
the supergravity backgrounds and the lightcone worldsheet theory on these 
backgrounds. In section \ref{section-D-branes}, we consider the BPS D-branes
on these backgrounds. We give the conditions to preserve the
supersymmetry by making use of the kappa symmetry projection on the
D-brane worldvolume.
Then we compare them to the results from the
analysis of the string worldsheet in the lightcone gauge.
In section \ref{section-general-backgrounds}, we study the D-branes in
more general cases by using the similar methods.
Section \ref{section-conclusion} is devoted to summary and discussion.


\section{Superstrings on supersymmetric pp-waves}
\label{section-review}

In this section, we construct some supersymmetric supergravity solutions
of pp-wave type and investigate the superstrings on these backgrounds 
in the lightcone gauge.
In the next subsection the supersymmetries on the pp-wave backgrounds
are examined and the worldsheet actions after taking the lightcone gauge are
proposed in subsection \ref{HVsol}.

\subsection{Supersymmetric solutions of type IIB pp-waves}

We consider the supersymmetric supergravity solutions of
pp-wave type constructed in \cite{Maldacena:2002fy}.
They are type IIB supergravity solutions with the following type of 
metric and non-trivial 5-form field strength $F_5$ as
\begin{align}
 ds^2 &= - 2dx^{+}dx^{-} + H(x^i) (dx^{+})^2 + d s^2_8 ~,\nn
 F_5 &= dx^+ \wedge \vp_4 (x^i)~,
\end{align}
where $x^{\pm}$ and $x^i$ are 2 longitudinal and 8 transverse
coordinates, respectively. 
The transverse space is assumed to be flat in order to make the
analysis simpler.
In this case, it is convenient to introduce the complex coordinates as 
$z^j = \frac{1}{\sqrt{2}} (x^j + i x^{j+4}),\ (j=1,2,3,4)$ and
the flat K\"aler metric $g_{i\jb}={\rm diag}(1,1,1,1)$.
Since the RR 5-form $F_5$ is self-dual, the 4-form $\vp_4$ has to be
anti-self dual in transverse 8-dimensions.  
The anti-self dual 4-form can be classified into two types.
They are (1,3) forms (and (3,1) forms) and (2,2) forms, which are denoted as
\begin{align}
 \vp_{mn}&:=\frac{1}{3!}\vp_{m\ib\jb\kb}\ve^{\ib\jb\kb\nb}g_{n\nb}~,\nn
 \vp_{l\bar{m}} &:= \frac{1}{2} g^{s \bar{s}} \vp_{l \bar{m} s \bar{s}} ~,
\end{align}
for the convenience.
The supersymmetries are generated by the Killing spinors
\begin{equation}
 \epsilon 
          = \epsilon_+ + \epsilon_- ~,\qquad
\epsilon_{+}:= - \frac{1}{2} \Gamma_+ \Gamma_- \epsilon~,\quad
\epsilon_{-}:= - \frac{1}{2} \Gamma_- \Gamma_+ \epsilon~, 
\end{equation}
which consist of 16 complex components.
The supersymmetries which are linearly realized
after taking the lightcone gauge are related to $\epsilon_+$,
therefore we will concentrate on these components.

The requirement of supersymmetry restricts the possible geometry.
If we require the (2,2) type of supersymmetry,
the allowed geometry is given by the metric and 4-forms%
\footnote{We use the holomorphic function $W$ different form the
one used in \cite{Maldacena:2002fy} by the factor $i$.}
\begin{align}
 & ds^2=-2dx^{+}dx^{-}-32(|\del_k W|^2 + |\vp_{j\bar{k}} z^j|^2)(dx^{+})^2 
   + 2 g_{i\ib}  dz^{i}d\zb^{\ib} ~,\nn
 &\vp_{mn}= i \del_m\del_n W~,~~
  \vp_{\bar{m} \bar{n}}= -i \del_{\bar{m}}\del_{\bar{n}} \bar{W}~,~~
  \vp_{l\bar{m}} = \mbox{(constant)} ~,
\label{22solution}
\end{align}
which are parametrised by a holomorphic function $W$ and a 
$4 \times 4$ hermitian traceless constant matrix $\vp_{j\bar{k}}$.
The Killing spinors are given by
\begin{align}
 \epsilon_{+}&= \alpha \vac + \zeta \vact~,\nn
 \epsilon_{-}&= 2 \Gamma_{-} 
 [ \zeta \del_{\kb} \bar W - i \alpha \vp_{j\bar{k}} z^j ] \Gamma^{\kb}\vac
             + 2 \Gamma_{-} 
 [ - \alpha \del_{k} W - i\zeta \vp_{k\jb} \bar{z}^{\jb} ] 
              \Gamma^{k} \vact~,
\label{22killing}
\end{align}
where $\alpha$ and $\zeta$ are constant parameters. 
The notation of Gamma matrices and the definition of vacua are
summarised in appendix A. 

Furthermore, there are solutions which preserve (1,1) type of supersymmetry.
The metric and the 4-form are given by
\begin{align}
 & ds^2=-2dx^{+}dx^{-}-32(|\del_k U|^2 )(dx^{+})^2 
   + 2 g_{i\ib}dz^{i}d\zb^{\ib} ~,\nn
 &\vp_{mn}=\del_m\del_n U~,~~
  \vp_{\bar{m} \bar{n}}=\del_{\bar{m}}\del_{\bar{n}} U~,~~
  \vp_{l\bar{m}} =  \del_{l}\del_{\bar{m}} U~,
\end{align}
where $U$ is a real harmonic function. 
The Killing spinors can be written as 
\begin{align}
 \epsilon_{+}&= - \zeta \vac + \zeta \vact~,\nn
 \epsilon_{-}&= 2i \Gamma_{-} \zeta \del_{\kb} U \Gamma^{\kb}\vac
             - 2i \Gamma_{-} \zeta \del_{k} U \Gamma^{k} \vact~.
 \label{11killing}
\end{align}

\subsection{The string actions in the lightcone gauge}
\label{HVsol}

The linearly realized supersymmetry on the worldsheet in the lightcone 
gauge $x^+=\tau$
is related to the spinor $\epsilon_+$, as we mentioned above. 
For the (2,2) supersymmetric solutions,
the action is given by
\begin{align}
 S &= \frac{1}{4\pi \alpha'} \int d \tau \int_0^{2\pi \alpha' |p_-|} d \sigma
 \left[L_{K}+L_{W}+L_{V}\right] ~,\nn
  L_K&=
      \int d^4 \theta
                              g_{i\jb}\Phi^{i}\Phb^{\jb}~, \qquad
 L_{W}=\frac12 \left(\int d^2 \theta W(\Phi)
 +(c.c.) \right)~,\nn
 L_{V}&=-|m|^2 g_{i\jb}V^{i}V^{\jb}
 -\frac{i}{2}(g_{i\ib}\del_{j}V^{i}-g_{j\jb}\del_{\ib}V^{\jb})
  (m\psb_{-}^{\ib}\psi_{+}^{j}+\mb\psb_{+}^{\ib}\psi_{-}^{j})~,
\label{LG}
\end{align}
where the chiral superfield $\Phi^i$ is expanded as
\begin{equation}
 \Phi^i = z^i + \sqrt{2} \ta^+ \psi^i_+ + \sqrt{2} \theta^{-} \psi^i_{-} 
 + 2 \ta^{+} \ta^{-} F^i + \cdots ~,
\end{equation}
and the vector $V$ is related to $\vp_{i\jb}$ as
\begin{equation}
 V_{i}=i \vp_{i\jb}\zb^{\jb}~,\quad V_{\jb}=-i \vp_{i\jb}z^{i}~.
\end{equation}
The indices are raised and lowered by $g^{i\jb}$ and $g_{i\jb}$, respectively.
Our convention of Landau-Ginzburg models are summarised in appendix \ref{22LG}.
The D-branes in the case with $V^{i}=0$ are considered in the next section
and the backgrounds with $V^{i}\ne 0$ 
are treated in subsection \ref{subsec-non-zero-V}.

The ${\cal N} = (1,1)$ supersymmetric action is of the form
\begin{equation}
 S = \frac{1}{2 \pi \alpha'} \int d \tau \int _0 ^{2\pi \alpha'|p_-|}
   d \sigma \int d^2 \theta \left(
        \frac12 D_{+} \Phi^{I} D_{-} \Phi^{I} + i U(\Phi)
   \right) ~,
 \label{N=1action}
\end{equation}
where the $N=1$ superfield $\Phi^{I}$ and supercovariant derivative
$D_{\pm}$ are defined as
\begin{equation}
 \Phi^{I}=x^{I}+\ta^{+}\psi_{+}^{I} 
 + \ta^{-}\psi_{-}^{I} +  \ta^{+}\ta^{-}F^I ~,
\qquad 
 D_{\pm}=\deldel{}{\ta^{\pm}}+i\ta^{\pm}\del_{\pm} ~.
\end{equation}
For the superstrings on pp-waves,
we have to restrict the superpotential to be harmonic. 
Our convention for the $\Ncal=(1,1)$ Landau-Ginzburg models is summarised
in appendix \ref{11LG}. We consider the D-branes in $\Ncal=(1,1)$ backgrounds
in subsection \ref{11D-brane}.


\section{D-branes in supersymmetric pp-waves}
\label{section-D-branes}

In this section we consider D-branes in the supersymmetric pp-waves
analysed in the previous section.
In the thin brane approximation, it is effective to use the worldvolume
approach.
The action on the $(p+1)$ dimensional worldvolume can be given by the sum of
the DBI and WZ actions as
\begin{equation}
 I_p = - T_{p} \int _M d^{p+1} \xi
  e^{-\phi} \sqrt{-\det(G_{ab} + {\cal F}_{ab})} 
       + T_{p} \int _M e^{\cal F} \wedge C ~, 
\end{equation}
where $T_{p}$ is the D$p$-brane tension expressed as 
$T_{p}=(2\pi)^{-p}(\alpha')^{-(p+1)/2}g_s^{-1}$.
We use $\xi^a$ $(a=0,\cdots,p+1)$ as the coordinate of the worldvolume and
$G_{ab}$ as the induced metric. We also define
${\cal F} = 2\pi \alpha' F-B$, where
$F$ is the field strength on the worldvolume and $B$ is the pullback of
the NSNS 2-form. The pullback of the RR gauge potentials is represented
as $C = \oplus_n C_n$.
For a while, we set ${\cal F} = 0$ and include this flux in subsection 
\ref{subsec-gauge}.

We are interested in the supersymmetric D-branes, since they are
expected to be stable.
In the supersymmetric D-brane configuration, we can define the kappa
symmetry projection as
 \cite{Cederwall:1997pv,Aganagic:1997pe,Cederwall:1997ri,Bergshoeff:1997tu,%
                   Aganagic:1997nn,Bergshoeff:1997kr} 
\begin{equation}
 \Gamma d^{p+1} \xi 
   = - e^{-\phi} {\cal L}^{-1}_{DBI}  e^{\cal F} \wedge X|_{vol}~,\qquad
 X = \oplus_n \Gamma_{(2n)} K^n I ~,
\label{projection}
\end{equation}
which satisfy $(\Gamma)^2 =1$.
The actions of $K$ and $I$ to the spinors are given by $K \psi = \psi^*$ and
$I \psi = -i \psi $, respectively and the Gamma matrices are defined by
\begin{equation}
 \Gamma_{(n)} = \frac{1}{n!} d \xi^{a_n} \wedge \cdots \wedge d\xi^{a_1} 
   \del_{a_1} x^{m_1} \cdots \del_{a_n} x^{m_n} \Gamma_{m_1 \cdots m_n} ~.
\end{equation}
The supersymmetries in the D-brane configuration are related to the
Killing spinors which satisfy
\begin{equation}
 \Gamma \epsilon = \epsilon ~,
\label{kappa}
\end{equation}
therefore the task we have to do is to look for the configuration
where the non-trivial Killing spinors satisfy (\ref{kappa}).

In this section we only consider the D-branes in the
(2,2) supersymmetric solutions only with non-zero superpotential $W$ as
\begin{eqnarray}
 &&ds^2=-2dx^{+}dx^{-}-32|\del_k W|^2(dx^{+})^2 + dz^{i}d\zb^{\ib}~,\nn
 &&\vp_{mn}= i \del_m\del_n W~,\qquad 
   \vp_{\bar{m}\bar{n}}= -i \del_{\bar{m}} \del_{\bar{n}} \bar{W} ~.
\end{eqnarray}
In the next section we will extend to the D-branes in more general
configurations. 

\subsection{D-brane wrapped on a complex submanifold}

We construct the D-branes tangent to the $x^{\pm}$ directions in order
to compare with the open string actions in the lightcone gauge.  
Therefore the simplest D-brane is the D1-brane with $\xi^{\pm} = x^{\pm}$,
where $\xi^{\pm}=\frac{1}{\sqrt{2}}(\xi^0 \pm \xi^1)$.
For this D-brane configuration, the kappa symmetry projection 
(\ref{projection}) can be written as
\begin{equation}
 \Gamma= - i\Gamma_{+-} K ~.
\end{equation}
By using the expression of the Killing spinors (\ref{22killing}),
we find
\begin{eqnarray}
 &&\Gamma \epsilon_{+}=  i\alpha^* \vact + i\zeta^* \vac~,\nn
 &&\Gamma \epsilon_{-}=- 2 i \zeta^* \Gamma_{-} \del_{k} W \Gamma^{k} \vact
      + 2 i \alpha^* \Gamma_{-} \del_{\kb} \bar W \Gamma^{\kb}\vac~.
\end{eqnarray}
Since the Killing spinors with
\begin{eqnarray}
 i\zeta^*=\alpha~,
\label{D0cond}
\end{eqnarray}
satisfy (\ref{kappa}) ($\Gamma \epsilon = \epsilon$),
we can conclude that this D1-brane is supersymmetric.

Let us turn to the D3-brane case.
The kappa symmetry projection can be defined as
\begin{equation}
 \hat{\Gamma} = 
 \Gamma d \xi^2 d \xi^3 =  \frac{- i}{\sqrt{- \det G}} 
  d Z^A d Z^B \Gamma_{+-} \Gamma_{AB} ~,
\end{equation}
where we use  
$Z^{A}$ $(A=1,2,3,4,\bar 1,\bar 2,\bar 3,\bar 4)$ with
$Z^{i}=z^{i},\ Z^{\ib}=\zb^{\ib}$.
Then we find
\begin{align}
 \hat{\Gamma} \epsilon_+ &= \frac{- i}{\sqrt{- \det G}} \bigl(
    \alpha d z^i d \bar{z}^{\ib} g_{i \ib} \vac 
   - \alpha d z^i d z^j \Gamma_i \Gamma_j \vac 
   - \zeta d z^i d \bar{z}^{\ib} g_{i \ib} \vact 
   - \zeta d \bar{z}^{\ib} d \bar{z}^{\jb} \Gamma_{\ib} \Gamma_{\jb} \vact 
   \bigr) \nn 
   &= (\alpha \vac + \zeta \vact )d \xi^2 d \xi^3 ~.
\end{align}
Thus we can see that the D3-brane is supersymmetric if
\begin{equation}
 \alpha = 0~,~~~ \omega = \sqrt{- \det G}d \xi^2 d \xi^3 ~,
\end{equation}
where we use the K\"{a}hler form $\omega=ig_{i\jb}dz^{i}d\zb^{\jb}$.
In the second condition, left-hand side is the pullback of the K\"ahler form,
and the right-hand side is the volume form of the D-brane world volume.
In other words, the supersymmetric D3-brane should be wrapped on a complex
submanifold.
  
{}From the above lessen, we use the holomorphic embedding $z^i(w)$ and 
$\bar{z}^{\ib}(\bar{w})$ with 
$w=\frac{1}{\sqrt{2}}(\xi^2 + i  \xi^3)$  and
$\bar{w}=\frac{1}{\sqrt{2}}(\xi^2 - i \xi^3)$.
In this case, the kappa symmetry projection is given by
\begin{equation}
 \Gamma:=- \frac{\del_{w}z^{i}\del_{\wb}\zb^{\ib}}{|\del_{w}z |^2}
   \Gamma_{+-}\Gamma_{i\ib}~,
\label{D3projection}
\end{equation}
and then we obtain
\begin{align}
 \Gamma \epsilon_{+}&= - \alpha \vac + \zeta \vact~,\nn
 \Gamma \epsilon_{-}&=  2 \zeta \Gamma_{-} \del_{\kb} 
    \bar W \Gamma^{\kb}\vac
      + 2 \alpha \Gamma_{-} \del_{k} W \Gamma^{k} \vact\nn
   &-\frac{4 \zeta}{|\del_w z|^2}\Gamma_{-}
    (\del_{\wb}\zb^{\kb}\del_{\kb}\bar W)(\del_{w}z^{i}\Gamma_{i})\vac
  - \frac{4 \alpha}{|\del_w z|^2}\Gamma_{-}
    (\del_{w}z^{k}\del_{k} W)(\del_{\wb}z^{\ib}\Gamma_{\ib})\vact~.
\label{D3e-}
\end{align}
Therefore we have non-trivial Killing spinors 
which satisfy $\Gamma \epsilon = \epsilon$ (\ref{kappa}) when
\begin{eqnarray}
\alpha=0~,\quad \del_{w}z^{k}\del_{k} W=0 ~,
\label{D3cond}
\end{eqnarray}
and in this case the D3-brane becomes supersymmetric.

The higher dimensional D-branes can be  analysed in the similar way and
we can see that the D-brane should be wrapped
on a complex submanifold\footnote{There are the other kind of
supersymmetric D-branes in the case of D5-branes and 
we investigate them in the next subsection.}.
Therefore we embed the D$(2n+1)$-brane $(n=2,3,4)$ 
in the holomorphic way as $z^i(w_1, \cdots, w_n)$ and 
$\bar{z}^{\ib} (\bar{w}_1 \cdots \bar{w}_n)$ with 
$w^a=\frac{1}{\sqrt{2}}(\xi^{2a} + i  \xi^{2a+1})$  and
$\bar{w}^{\bar{a}}=\frac{1}{\sqrt{2}}(\xi^{2a} - i \xi^{2a+1})$.
We also denote the determinant of the induced metric as
\begin{align}
  h:=\det \left[\deldel{z^{i}}{\xi^{a}} 
        \deldel{\zb^{\jb}}{\xi^{b}} g_{i \jb}+(a\leftrightarrow b)\right]
    =\left|\det\left[\deldel{z^{i}}{w^{p}} 
        \deldel{\zb^{\jb}}{\wb^{\bar{q}}} g_{i \jb}\right]\right|^2~,
\end{align}
where we use 
 $a,b=2,\dots,2n+1$, $p=1,\dots,n$ and  $\bar{q}=\bar{1},\dots,\bar{n}$.
In these cases, the kappa symmetry projections (\ref{projection}) are
given by  
\begin{align}
 \Gamma&= i h^{-1/2}
  {\del_{w_1}z^{i}\del_{\wb_1}\zb^{\ib}
        \del_{w_2}z^{j}\del_{\wb_2}\zb^{\jb}}%
       \Gamma_{+-}\Gamma_{i \ib j \jb} K ~~(\mbox{D5})~,\nn
 \Gamma&=h^{-1/2}
  {\del_{w_1}z^{i}\del_{\wb_1}\zb^{\ib}
        \del_{w_2}z^{j}\del_{\wb_2}\zb^{\jb}
        \del_{w_3}z^{k}\del_{\wb_3}\zb^{\kb}}%
       \Gamma_{+-}\Gamma_{i \ib j \jb k \kb} 
      ~~(\mbox{D7})~,\nn
 \Gamma&= -i h^{-1/2}
  {\del_{w_1}z^{i}\del_{\wb_1}\zb^{\ib}
        \del_{w_2}z^{j}\del_{\wb_2}\zb^{\jb}
        \del_{w_3}z^{k}\del_{\wb_3}\zb^{\kb}
        \del_{w_4}z^{\ell}\del_{\wb_4}\zb^{\lb}}%
       \Gamma_{+-}\Gamma_{i\ib j\jb k\kb \ell\lb}K
     ~~(\mbox{D9})~,
\end{align}
and the conditions that the Killing spinors satisfying (\ref{kappa})
($\Gamma \epsilon = \epsilon$) exist are
\begin{align}
 &- i\zeta^{*}=\alpha~, 
  ~~ \del_{w_1}z^{k}\del_{k} W=\del_{w_2}z^{k}\del_{k} W=0 ~~(D5)~,\nn
 &\zeta=0~, ~~ \del_{w_1}z^{k}\del_{k} W=
            \del_{w_2}z^{k}\del_{k} W=\del_{w_3}z^{k}\del_{k} W=0~~(D7)~,\nn
 & i\zeta^{*}=\alpha~, 
  ~~ \del_{w_1}z^{k}\del_{k} W=\del_{w_2}z^{k}\del_{k} W=
             \del_{w_3}z^{k}\del_{k}W=\del_{w_4}z^{k}\del_{k} W=0 ~~(D9)~.
\label{Dpcond}
\end{align}

In summary, we have shown that the D-branes are supersymmetric if they
are wrapped on complex submanifolds and satisfy
\begin{equation}
 \del_{a} W = 0 ~,~~  \del_{a} \bar W = 0 ~,~~ 
 \eta_{+} = e^{i\theta} \eta_{-} ~,
\label{holomorphic}
\end{equation}
where $a$ represents the tangent direction of the branes and 
the phase $e^{i\theta}$
is determined for D$p$-brane as $e^{i\theta}=(-i)^{(p-1)/2}$. 
Here we also define as
\begin{equation}
 \eta_{+} := \alpha + \zeta^* ~,~~ \eta_{-} := - i\alpha + i\zeta^* ~,
  ~~ \bar{\eta}_{\pm} = (\eta_{\pm})^* ~.
\label{eta}
\end{equation}
The corresponding string worldsheet in the lightcone gauge is given by
(\ref{LG}) replaced by the closed worldsheet with the open worldsheet
(${\mathbb R} \times [0,\pi \alpha' |p_-|]$).
Supersymmetric boundary condition on this worldsheet was investigated in
\cite{Hori:2000ck} and our
results (\ref{holomorphic}) correspond to the conditions for the B-branes.
{}From this reason, we also call the D-branes constructed in this subsection
as B-type D-branes.

\subsection{D-brane wrapped on a special Lagrangian submanifold}

For the D5-brane, we can construct non-holomorphic type of 
supersymmetric D-brane, which preserves the supersymmetry of the
type different from  (\ref{holomorphic}).
We use the coordinate of the worldvolume as 
$(\xi^{+},\xi^{-},\xi^{2},\dots,\xi^{5})$ and the coordinate of the
spacetime transverse space as $Z^{A}$
$(A=1,2,3,4,\bar 1,\bar 2,\bar 3,\bar 4)$ with
$Z^{i}=z^{i},\ Z^{\ib}=\zb^{\ib}$.
Moreover, we use the K\"ahler form as $\omega=ig_{i\jb}dz^{i}d\zb^{\jb}$
and the holomorphic 4-form as $\Omega=4dz^{1}dz^{2}dz^{3}dz^{4}$, which
satisfy $2^4 \omega^4=4!\Omega\Omegab$.

For the kappa symmetry projection (\ref{projection}), it is convenient
to use
\begin{equation}
 \hat{\Gamma}:=\Gamma d\xi^2 d\xi^3 d\xi^4 d\xi^5
        = \frac{- i}{\sqrt{h}}\frac{1}{4!}dZ^{A}dZ^{B}dZ^{C}dZ^{D}
       \Gamma_{+-ABCD}K ~,
\end{equation}
where $h$ is the determinant of the induced metric defined as
\begin{align}
   h:=\det \left[\deldel{z^{i}}{\xi^{a}} 
        \deldel{\zb^{\jb}}{\xi^{b}} g_{i \jb}+(a\leftrightarrow
 b)\right] ~,\qquad a,b=2,3,4,5,
\end{align}
and the differential forms are considered to be pulled back to the
D-brane worldvolume.

In this notation, the actions to the $\epsilon_+$ Killing spinors are
given by
\begin{align}
 &\hat{\Gamma} \epsilon_{+}=\frac{i}{\sqrt{h}} \Bigg(
     4\alpha^* d\zb^{\bar 1}d\zb^{\bar 2}d\zb^{\bar 3}d\zb^{\bar 4}\vac
    +4\zeta^* dz^{1}dz^{2}dz^{3}dz^{4}\vact 
     +\frac13\alpha^*dz^{i}d\zb^{\jb}d\zb^{\kb}d\zb^{\lb}g_{i\jb}
                                                     \Gamma_{\kb\lb}\vact
   \nn & \qquad
   +\frac13\zeta^*d\zb^{\ib}dz^{j}dz^{k}dz^{\ell}g_{\ib j}
                                                     \Gamma_{k \ell}\vac 
     -\frac12 \alpha^{*}
    dz^{i}dz^{j}d\zb^{\kb}d\zb^{\lb}g_{i\kb}g_{j\lb}\vact
      -\frac12 \zeta^{*}d\zb^{\ib}d\zb^{\jb}    
                        dz^{k}dz^{\ell}g_{\ib k}g_{\jb \ell}\vac
  \Bigg)\nn
  & ~ =\frac{i}{\sqrt{h}} \Bigg(
     \alpha^* \Omegab\vac + \zeta^* \Omega\vact 
     +\frac{\alpha^*}{3i}\omega d\zb^{\kb}d\zb^{\lb} \Gamma_{\kb\lb}\vact
-\frac{\zeta^*}{3i}\omega dz^{k}dz^{\ell}       
     \Gamma_{k \ell}\vac
     -\frac{\alpha^{*}}{2}\omega^2\vact-\frac{\zeta^{*}}{2}\omega^2\vac
  \Bigg) .
\end{align}
Here we are looking  for the Killing spinors 
different from (\ref{holomorphic}), say\footnote{
The phase factor can be set to one by redefining the complex coordinates.
}
\begin{equation}
 \eta_{+}=\etb_{-} ~~ \mbox{or} ~~ i\alpha^{*}=\alpha~,\ i\zeta^{*}=\zeta ~.
\end{equation}
By assigning this condition, we obtain the constraints
\begin{eqnarray}
 \omega=0 ~,\qquad \Omega=\sqrt{h}\;
   d\xi^{2}d\xi^{3}d\xi^{4}d\xi^{5}
\end{eqnarray}
on the D-brane worldvolume.
Therefore the D-brane should be wrapped on 
a special Lagrangian submanifold $\gamma$, which is defined by
\begin{eqnarray}
 \omega|_{\gamma}=0~,\qquad \im \Omega|_{\gamma}=0 ~.
\label{SL} 
\end{eqnarray}

The actions to the $\epsilon_-$ part of the Killing spinors
are similarly obtained as
\begin{eqnarray}
 \lefteqn{\hat{\Gamma} 2 \zeta \Gamma_{-}\del_{\mb}\bar W\Gamma^{\mb}\vac
=  \frac{- 2 i \zeta^{*}}{\sqrt{h}} \Gamma_- \del_{m}W }\nn
&& \times\Bigg(
\frac13 dz^{i}d\zb^{\jb}d\zb^{\kb}d\zb^{\lb}g_{i\jb}g^{m\mb}\Gamma^{\nb}
               \ve_{\nb\mb\kb\lb}\vac
 -\frac23 dz^{i}d\zb^{\jb}d\zb^{\kb}d\zb^{\lb}\delta_{i}^{m}\Gamma^{\nb}
               \ve_{\nb\jb\kb\lb}\vac\nn
 &&\qquad
 -\frac12 dz^{i}dz^{j}d\zb^{\kb}d\zb^{\lb}g_{i\kb}g_{j\lb}\Gamma^{m}\vact
  - 2 dz^{i}dz^{j}d\zb^{\kb}d\zb^{\lb}\delta_{i}^{m}g_{j\kb}\Gamma_{\lb}\vact
\Bigg) ~.
\label{SLe-}
\end{eqnarray}
The other parts are obtained by the complex conjugation
 and exchanging $\alpha$ and $\zeta$ of the above equation.
By taking account of $\omega=0$ and $\im \Omega=0$, the condition of 
$i\zeta^{*}=\zeta$ leads to
\begin{eqnarray}
 \frac{4}{3} \frac{1}{\sqrt{h}}
\del_{m}W dz^{m}d\zb^{\jb}d\zb^{\kb}d\zb^{\lb}\ve_{\mb\jb\kb\lb} 
     = 2\del_{\mb}\bar W d\xi^{2}d\xi^{3}d\xi^{4}d\xi^{5}~,
\end{eqnarray}
hence
\begin{eqnarray}
 \frac{1}{3!} dW d\zb^{\jb}d\zb^{\kb}d\zb^{\lb}\ve_{\mb\jb\kb\lb}
  =\del_{\mb}\bar W d\zb^{1}d\zb^{2}d\zb^{3}d\zb^{4}~,
\end{eqnarray}
and this condition is equivalent to 
\begin{eqnarray}
 dW = d \bar W.
\label{SL2}
\end{eqnarray}

This type of D-brane corresponds to the A-brane in the terms of
\cite{Hori:2000ck} when we consider the open string worldsheet in the
lightcone gauge, thus we also call this brane as A-brane.
The condition (\ref{SL2}) can be reproduced by the analysis of
\cite{Hori:2000ck}, 
however the condition (\ref{SL}) is slightly different.
Our Killing spinor analysis has shown that the supersymmetric D-brane of
this type should be wrapped on a {\em special} Lagrangian submanifold, on the
other hand, the
analysis of \cite{Hori:2000ck} gives only the requirement that the
A-brane should be wrapped on a Lagrangian
(not necessarily special Lagrangian) submanifold ($\omega=0$).
Some comments on this point are given in section
\ref{section-conclusion}.

\subsection{Examples}

In this subsection, we show some examples of the 
supersymmetric D-brane configurations
considered in the above subsections. In particular, we consider the 
maximally supersymmetric case $ W= -i \sum_{j}(z^{j})^2$ and compare it to the
known results.

First, let us comment on D9-branes. Since the superpotential should be
constant on the supersymmetric D-brane worldvolume, D9-brane cannot
be supersymmetric for the nontrivial superpotential $W$.

Secondly, let us go to D7-branes. For a nontrivial superpotential $W(z)$,
the B-type D7-brane worldvolume should be identical to a hyper surface 
$ i W(z)=c,\ (c:$ constant). For example, in the maximally supersymmetric case
($W=-i\sum_{j}(z^{j})^2$), the D7-brane is expressed as
\begin{align}
 (z^1)^2+(z^2)^2+(z^3)^2+(z^4)^2=c ~.
\end{align}
This surface has the same topology and
complex structure as a (deformed) conifold.
Note that the flat D7-brane expressed as $(+,-,4,2)$ in \cite{Skenderis:2002vf}
is not a B-type brane in our terms. The $(+,-,4,2)$ brane does not preserve
the supersymmetry expressed by the Killing spinor of the type (\ref{22killing}).
In the maximally supersymmetric plane wave case, 
there are many Killing spinors besides ones expressed as eq. (\ref{22killing}). The $(+,-,4,2)$ brane preserves
nontrivial linear combinations of these extra Killing spinors
and ones of (\ref{22killing}).

Thirdly, we consider the B-type D5-branes and D3-branes. These branes can
take the various shapes. For the maximally supersymmetric case,
there is a flat D5-brane expressed as
\begin{align}
z_1=i z_2 ~,\quad z_3=i z_4 ~,
\end{align}
and a flat D3-brane expressed as
\begin{align}
z_1=i z_2 ~,\quad z_3=a ~,\quad  z_4=b ~, \qquad(a,b:\text{constants})~.
\end{align}
Note that these branes are not the ones classified in \cite{Skenderis:2002vf}.
These branes are extended to oblique directions and cannot be expressed as
$(+,-,m,n)$. The $(+,-,3,1)$ and $(+,-,2,0)$ branes are not B-type D-branes
in our terms, for the same reason as the $(+,-,4,2)$ D7-brane.

Fourthly, let us turn to D1-branes. For the maximally supersymmetric case,
this brane is the same as $(+,-,0,0)$ in \cite{Skenderis:2002vf}.

Finally, we comment on the A-type D5-branes. A typical example of
special Lagrangian submanifold 
is the worldvolume of $(+,-,4,0)$ brane in \cite{Skenderis:2002vf}.
As discussed in \cite{Skenderis:2002vf}, this brane is not a solution of
the equation of motion without worldvolume gauge field excitation.
Moreover, this brane does not satisfy the condition
of superpotential ($\im W=$ constant) obtained in this section. 
We will discuss the gauge field
and equation of motion in subsection \ref{subsec-gauge}. If we include
the gauge field, the condition of superpotential is modified.
As a result, the $(+,-,4,0)$ brane with appropriate gauge field
excitation {\em is} an A-type D-brane in our terms.


\section{D-branes in more general cases}
\label{section-general-backgrounds}

\subsection{Inclusion of gauge field excitations}
\label{subsec-gauge}

In the previous section, we have considered the D-brane configurations
without gauge fields. 
Here we include the gauge fields of the type
\begin{equation}
 F = \sum_{a} F_{+a} d \xi^+ \wedge d \xi^a ~,~~
 F_{+a} = \partial_a A_+ (\xi) ~, ~~ (a=1,\cdots,p-1) ~.
\end{equation}
In this case, the equation of motion of $x^{I}$ is given in 
\cite{Skenderis:2002vf}.
Let us define
\begin{eqnarray}
 &&M_{a'b'} = \partial_{a'} x^I \partial_{b'} x^J g_{IJ} 
               + 2 \pi \alpha'  F_{a'b'} ~,~~
 (a',b',c'=\pm,1,\cdots,p-1) ~, \nonumber \\
 &&M^{a'b'} M_{b'c'} = \delta^{a'}_{~c'} ~,~~
   G^{a'b'} = M^{(a'b')} ~,~~ \theta^{a'b'} = M^{[a'b']}~, 
\end{eqnarray} 
then the equation of motion can be written as
\begin{equation}
 \partial_{a'} (\sqrt{-M} G^{a'b'} \partial_{b'} x^I  ) = 0 ~.
\end{equation}
This equation does not give more constraints to the coordinates of the 
D-brane wrapped on a complex submanifold or a special Lagrangian
submanifold. 

On the other hand, the equation of motion of the gauge field $A^i$
may give some constraints. 
For a B-type brane, it is given as
\begin{equation}
 \partial_{a'} (\sqrt{-M} \theta^{a'b'}) = 0 ~,
\end{equation}
hence we can see that the configuration without gauge field excitation
 satisfies
the equation of motion as well as the one with some solutions $F_{+i}$. 
However, for D-branes wrapped on special Lagrangian submanifolds, we find
\begin{equation}
 \partial_{a'} (\sqrt{-M} \theta^{a'-}) = 
 \frac{1}{4!} \epsilon^{-+ijkl} F_{+ijkl} ~,
 ~~ \partial_{a'} (\sqrt{-M} \theta^{a'a}) = 0 ~, 
\end{equation}
because there may be contributions from the WZ action in this case.
If the condition $\im W=$(constant) is satisfied, the right-hand side of the
first equation vanishes. On the other hand, even if $\im W=$(constant) is not
satisfied, we may obtain a BPS D-brane by introducing the gauge field $F$.

First, let us examine the B-type D3-brane. The other dimensional B-branes
can be analysed in a similar way.
In this case, the kappa symmetry projection (\ref{projection}) is
given by 
\begin{equation}
\Gamma = \Gamma_{(1)} + \Gamma_{(2)} ~,
\end{equation}
where $\Gamma_{(1)}$ is the previous one (\ref{D3projection})
and $\Gamma_{(2)}$ is the term added additionally as
\begin{equation}
 \Gamma_{(2)} =  \frac{\partial_{\wb} \zb^{\ib}}
               {|\partial _w z|^2} F_{+ w} \Gamma_- \Gamma_{\ib}K
        - \frac{\partial_w z^i}
               {|\partial _w z|^2} F_{+\wb} \Gamma_- \Gamma_i K ~. 
\end{equation}
The action of $\Gamma_{(1)}$ to the Killing spinor is the same as before
and the one of $\Gamma_{(2)}$ is given as
\begin{equation}
 \Gamma_{(2)} \epsilon  = 
  \alpha^* \frac{\partial_{\wb} \zb^{\ib}}
               {|\partial _w z|^2} F_{+ w} \Gamma_- \Gamma_{\ib} \vac
 - \zeta^* \frac{\partial_w z^i}
               {|\partial _w z|^2} F_{+\wb} \Gamma_- \Gamma_i \vact ~.
\end{equation}
By adding to the action of $\Gamma_{(1)}$ (\ref{D3e-}),   
we can see that the conditions (\ref{D3cond}) are replaced by\footnote{
We can use $\zeta = ie^{2i\beta}\zeta ^*$ for the parameter of the
Killing spinor. In that case, the condition of the superpotential is
slightly modified by the phase factor.
}
\begin{equation}
 \alpha = 0 ~, ~~\zeta =  i \zeta^* ~,~~
 \partial_w z^k \partial_k W +  \frac{i}{4} F_{+w} =0 ~. 
\label{modD3}
\end{equation}
Therefore, if we include the non-trivial gauge fields, we can only 
construct the B-type D-branes which preserve at most $1/4$ supersymmetry.

Next, let us consider the A-type D5-brane.
Including non-trivial field strength $F_{+a}$, the kappa symmetry
projection (\ref{projection}) becomes
$\hat{\Gamma} = \hat{\Gamma}_{(1)} + \hat{\Gamma}_{(2)}$ with the
additional term $\Gamma_{(2)}$ as
\begin{equation}
 \hat{\Gamma}_{(2)} = 
  \frac{i}{\sqrt{h}}  F_{+a} d \xi^a \frac{1}{3!}
 d Z^A d Z^B d Z^C \Gamma_- \Gamma_{ABC} ~.
\end{equation}
Here we should notice that the condition (\ref{kappa}) can be separated as
\begin{equation}
 \hat{\Gamma}_{(1)} \epsilon_+ = \epsilon_+ d^4\xi ~,~~
 \hat{\Gamma}_{(1)} \epsilon_- + \hat{\Gamma}_{(2)} \epsilon_+ 
  = \epsilon_-d^4\xi
   ~.
\end{equation}
The first equation implies that 
the supersymmetric A-brane must be wrapped on a special Lagrangian
submanifold.
Using $\omega|_{\gamma}=0$, we find
\begin{align}
\hat{\Gamma}_{(2)} \epsilon_+ &= \frac{- 2 i \alpha}{\sqrt{h}} 
      F_{+a}d \xi^a \frac{1}{3!} 
     \epsilon_{jklm} d z^j d z^k d z^l \Gamma_- \Gamma^m \vact
 \nonumber \\ &
+  \frac{- 2 i \zeta}{\sqrt{h}} 
      F_{+a}d \xi^a \frac{1}{3!} 
     \epsilon_{\jb \kb \lb \mb} d z^{\jb} d z^{\kb} d z^{\lb} 
    \Gamma_- \Gamma^{\mb} \vac ~.
\end{align}
This equation and (\ref{SLe-}) imply 
\begin{equation}
 \frac{1}{3!} \left( dW + \frac{i}{4 } F_{+a} d \xi^a \right) 
 d\zb^{\jb}d\zb^{\kb}d\zb^{\lb}\ve_{\mb\jb\kb\lb}
  =\del_{\mb}\bar W d\zb^{1}d\zb^{2}d\zb^{3}d\zb^{4}~,
\end{equation}
where we use $ i \zeta^* = \zeta$.
Thus the A-type D5-brane  preserves 
supersymmetry if the superpotential satisfy
\begin{equation}
 \partial_a (W -  \bar{W}) + \frac{i}{4} F_{+a} = 0~,
\label{modAD5}
\end{equation}
for the tangent direction $a$ of the brane.

The above Killing spinor results can be reproduced by analysing the  
open string worldsheet in the lightcone gauge.
The inclusion of the gauge fields corresponds to the addition of the following
boundary potential;
\begin{equation}
 S_B = 
 \frac{1}{2} \int _{\partial \Sigma} d \tau Y ( z^i, \zb^{\ib})~.
\end{equation}
This action is invariant under the transformation $\delta_B z^i = 0 $
and $\delta_B \psi_{B}^i = \kappa f^i (z, \zb) $, where $\kappa$ is a
spinor and $\psi_B^i$ is the fermionic coordinate at the boundary.
The supersymmetry transformation at the boundary can be modified by this
transformation.
By taking the variation of the action by this transformation, 
we obtain the boundary conditions of the fields.

For the A-type boundary condition $\eta_+ = \bar{\eta}_-$,
what we have to do is only replacing 
$\partial_m W$ with $\partial_m W + 2 i \partial_m Y$, then the boundary
condition for superpotential is modified as \cite{Lindstrom:2002jb}
\begin{equation}
 \partial_a (W -\bar{W}) + 2 i \partial_a Y = 0 ~,
\end{equation} 
which is the same as (\ref{modAD5}).
For the B-type boundary condition $\eta_+ = \eta_-$, 
we can only preserve 1/4 supersymmetry if we include non-zero gauge fields.
In this case, we have to replace 
$\partial_m W$ with $\partial_m W + 2 i \partial_m Y$,
where we assign $\eta_+ = \bar{\eta}_+ $.
Therefore the boundary conditions becomes
\begin{equation}
 \partial_a W + 2 i \partial_a Y = 0 ~, 
\end{equation}
which corresponds to the Killing spinor results (\ref{modD3}).

\subsection{The background with non-zero (2,2)-form}
\label{subsec-non-zero-V}

Let us consider the background with non-zero $(2,2)$-form $\vp_{m\nb}$.
In this case, we introduce a harmonic function 
$U=\vp_{m\nb}z^{m}\zb^{\nb}$, and write the Killing spinor as
\begin{align}
 &\epsilon_{+}=\alpha \vac+\zeta \vact~,\qquad
 \epsilon_{-}=\beta_{\mb}\Gamma_{-}\Gamma^{\mb}\vac
                        +\delta_{m}\Gamma_{-}\Gamma^{m}\vact~,\nn
 &\beta_{\mb}:=2\del_{\mb}(- i \alpha U + \zeta \bar W)~,\qquad
 \delta_{m}:=2\del_{m}(- i \zeta U - \alpha W) ~.
\end{align}
In the analysis of the kappa symmetry projection,
 $ \Gamma \epsilon_{+}$ is the same as the previous section, and 
the conditions obtained from $\Gamma \epsilon_{+}=\epsilon_{+}$
should hold also in this case. We see below the additional conditions
derived from $\Gamma \epsilon_{-}=\epsilon_{-}$.

For a D0-brane, $\Gamma\epsilon_{-}$ becomes
\begin{align}
 &\Gamma=-i\Gamma_{+-}K~,\quad
 \Gamma\epsilon_{-}=-i\beta_{m}\Gamma_{-}\Gamma^{m}\vact
                   -i\delta_{\mb}\Gamma_{-}\Gamma^{\mb}\vac~,
\end{align}
where we introduce
\begin{align}
 &\beta_{m}:=(\beta_{\mb})^*=2\del_{m}(i\alpha^* U + \zeta^*  W)~,~~
 \delta_{\mb}:=(\delta_{m})^*=2\del_{\mb}(i\zeta^* U - \alpha^* \bar W)~.
\end{align}
As a result, $\Gamma \epsilon_{-}=\epsilon_{-}$ implies
\begin{align}
 i \zeta^* =\alpha ~.
\end{align}
There is no additional condition for a D0-brane.

In contrast, for a B-type D3-brane, $\Gamma \epsilon_{-}=\epsilon_{-}$ reads
$\beta_{\kb}=0$. From $\Gamma \epsilon_{-}=\epsilon_{-}$ we obtain
$\alpha=0$. 
Consequently, the kappa symmetry projection implies $\del_{\kb}U=0$,
and there is no B-type D3-brane for non-zero $\vp_{m\nb}$.

Let us turn to the B-type D5-brane. In this case,
$\Gamma \epsilon=\epsilon$ reads 
\begin{align}
 &\zeta^{*}=i\alpha~,\qquad \del_{w^{a}}z^{k}\beta_{k}=0~,\quad 
 \del_{\wb^{a}}\zb^{\kb}\delta_{\kb}=0~,\quad (a=1,2)~.
\end{align}
Therefore, for the supersymmetric B-type D5-brane, $U$ must be a constant on
the D5-brane worldvolume, in addition to the condition $W$ must be a
constant.

As a same manner, we analyse B-type D7-branes and D9-branes.
For a non-zero $\vp_{m\nb}$, B-type D7-branes do not exist for the same reason
as D3-branes. On the other hand,
 D9-branes also do not exist since $U$ cannot be constant
on the D9-brane worldvolume for a non-zero $\vp_{m\nb}$.

Finally, we examine the A-type D5-brane. We obtained from the kappa symmetry
analysis the following conditions on superpotential $W$ and real
harmonic function $U$ as
\begin{align}
 dW=d\Wb~,\qquad \del_{m}Udz^{m}+ \del_{\mb}U d\zb^{\mb}=0~,
\end{align}
where the differential forms are pulled back to the worldvolume.
The second condition shows that $U$ must be constant on the
D-brane worldvolume.

Now, let us turn to the worldsheet analysis of the Landau-Ginzburg
models. The variation of the action (\ref{LG}) becomes (see
eq.(\ref{22variation})) 
\begin{align}
 \delta S &= \int_{\del \Sigma}d\tau\Bigg\{
\frac12 g_{i\jb}\Big[
  -\eta_{+}\del_{+}\phb^{\jb}\psi^{i}_{-}
  -\eta_{-}\del_{-}\phb^{\jb}\psi^{i}_{+}
  +\etb_{+}\del_{+}\pha^{i}\psb^{\jb}_{-}
  +\etb_{-}\del_{-}\pha^{i}\psb^{\jb}_{+}
  \Big]\nn
 &\quad -\frac{i}{4}\Big[
     -\eta_{+}\del_{\ib}\Wb\psb^{\ib}_{+}
     +\eta_{-}\del_{\ib}\Wb\psb^{\ib}_{-}
     -\etb_{+}\del_{ i } W \psi^{ i }_{+}
     +\etb_{-}\del_{ i } W \psi^{ i }_{-}
    \Big]\nn
 &\quad -\frac12\Big[
 -\eta_{+} m  V_{ i }\psi^{ i }_{+}
 -\eta_{-}\mb V_{ i }\psi^{ i }_{-}
 +\etb_{+}\mb V_{\jb}\psb^{\jb}_{+}
 +\etb_{-} m  V_{\jb}\psb^{\jb}_{-}
   \Big] \Bigg\} ~.
 \label{22variation-boundary}
\end{align}

First, we consider the B-type supersymmetry $\eta_{+}=-\eta_{-}$.
In this case, eq.(\ref{22variation-boundary}) becomes
\begin{align}
 \delta S &= \int_{\del \Sigma}d\tau\Bigg\{
\frac12 g_{i\jb}\Big[
  -\eta_{+}(
       \del_{\tau}  \zb^{\jb}\{\psi_{-}^{i}-\psi_{+}^{i}\}
     + \del_{\sigma}\zb^{\jb}\{\psi_{-}^{i}+\psi_{+}^{i}\}
       )
  +\bar{\eta}_{+}(
       \del_{\tau}  z^{i}\{\psb_{-}^{\jb}-\psb_{+}^{\jb}\}
\nn & \qquad \qquad
     + \del_{\sigma}z^{i}\{\psb_{-}^{\jb}+\psb_{+}^{\jb}\}
       )
  \Big]
-\frac{i}{4}\Big[
     -\eta_{+}\del_{\ib}\Wb(\psb^{\ib}_{+}+\psb^{\ib}_{-})
     -\etb_{+}\del_{ i } W (\psi^{ i }_{+}+\psi^{ i }_{-})
    \Big]
\nn & \qquad \qquad
 -\frac12\Big[
 -\eta_{+}(m  V_{ i }\psi^{ i }_{+}-\mb V_{i}\psi^{i}_{-})
 +\etb_{+}(m  V_{\ib}\psb^{\ib}_{+}-\mb V_{\ib}\psb^{\ib}_{-})
   \Big] \Bigg\}~.
\end{align}
If we assume $m=-\mb$, the result is the same as the Killing spinor analysis.
In this case, for a Neumann direction $I$,
we have to set $\psi^{I}_{+}=\psi^{I}_{-}$, and also $\del_{I}W=V_{I}=0$. 
For a Dirichlet direction, $\psi^{I}_{+}= - \psi^{I}_{-}$ has to be
satisfied, and there is no more condition on $W$ and $V$. 
As a result, for the B-type D-brane,
$W$ and $U$ should be constant on the D-brane worldvolume.

Next, we consider the A-type boundary condition. 
If we set $\eta_{+}=\etb_{-}$,
the variation (\ref{22variation-boundary}) becomes
\begin{align}
 &\delta S = \int_{\del \Sigma}d\tau\Bigg\{
\frac12 g_{i\jb}\eta_{1}\Big[
-\del_{0}\zb^{\jb}(\psi_{-}^{i}+\psi_{+}^{i})
+\del_{0} z^{ i }(\psb_{-}^{\jb}+\psb_{+}^{\jb})
-\del_{1}\zb^{\jb}(\psi_{-}^{i}-\psi_{+}^{i})
+\del_{1} z^{ i }(\psb_{-}^{\jb}-\psb_{+}^{\jb})
\Big]
\nn&
+\frac12 g_{i\jb}i\eta_{2}\Big[
-\del_{0}\zb^{\jb}(\psi_{-}^{i}-\psi_{+}^{i})
-\del_{0} z^{ i }(\psb_{-}^{\jb}-\psb_{+}^{\jb})
-\del_{1}\zb^{\jb}(\psi_{-}^{i}+\psi_{+}^{i})
-\del_{1} z^{ i }(\psb_{-}^{\jb}+\psb_{+}^{\jb})
\Big]
\nn&
   -\frac{i}{4}\eta_{1}\Big[
      \del_{\ib}\Wb(\psb_{-}^{\ib}-\psb_{+}^{\ib})
     +\del_{i} W (\psi_{-}^{i}-\psi_{+}^{i})
   \Big]
   +\frac{1}{4}\eta_{2}\Big[
     -\del_{\ib}\Wb(\psb_{-}^{\ib}+\psb_{+}^{\ib})
     +\del_{i} W (\psi_{-}^{i}+\psi_{+}^{i})
   \Big] 
\nn&
  -\frac12\eta_{1} m \Big[
    V_{i}(\psi_{-}^{i}-\psi_{+}^{i})
    +V_{\ib}(\psb_{-}^{\ib}-\psb_{+}^{\ib})
   \Big]
  +\frac{i}{2}\eta_{2} m \Big[
    V_{i}(\psi_{-}^{i}+\psi_{+}^{i})
   -V_{\ib}(\psb_{-}^{\ib}+\psb_{+}^{\ib})
   \Big]
 \Bigg\}~,
\end{align}
where we use $\eta_{+}=\eta_1+i\eta_2,\ (\eta_1,\eta_2:$ real) and
assume $m=-\mb$. If we take it into account that
both $i\del_{0}z^{i}$ and $\del_{1}z^{i}$ are the holomorphic components of
normal vectors of the D-brane worldvolume,
we can read from the first and second line, that
both $u^{i}=\eta(\psi_{-}^{i}-\psi_{+}^{i})$
and $v^{i}=i\eta(\psi_{-}^{i}+\psi_{+}^{i})$
are the holomorphic components of tangent vectors of the worldvolume for a
real fermionic parameter $\eta$.
Therefore, we obtain the conditions that $\delta S$ vanishes as
\begin{align}
 &v^{I}\del_{I} (W-\Wb)=u^{I}\del_{I} (W-\Wb)=0 ~, \nn
 &v^{I} \del_{I} U = u^{I} \del_{I} U =0~,
\end{align}
where we use $V_{j}=i\del_{j} U$. This condition implies that
both $\im W$ and $U$ must be constant on the D-brane worldvolume.
This is the same result as the one obtained from the Killing spinor 
analysis\footnote{The discrepancy of ``special Lagrangian or
Lagrangian'' is still present.}.

Let us here comment on the phase of $m$. The phase of $m$ does not appear in
the supergravity solution (\ref{22solution})%
\footnote{The absolute value and the sign of $m$ can be absorbed 
into the hermitian matrix $\vp_{i\jb}$.}.
Assuming that $m=-\mb$, the constraints on
both A-type and B-type branes from worldsheet analysis
are consistent with the ones from spacetime analysis. 
For this reason, we claim that $m$
is a pure imaginary number in the worldsheet theory on the supergravity
background (\ref{22solution}).

\subsection{$(1,1)$ supersymmetric background}
\label{11D-brane}

In this section, we consider the D-branes in $(1,1)$ supersymmetric
background. We first consider the D-branes from the point of view of the
worldsheet Landau-Ginzburg theory. Then we compare it to the Killing spinor 
analysis. 

The superfield formalism of the $\Ncal=(1,1)$ Landau-Ginzburg theory is 
summarised in appendix \ref{11LG}.
The variation of the action (\ref{N=1action}) on the worldsheet with boundary
by the supersymmetry transformation can be calculated by using 
Eq.(\ref{11variation}) as (omitting the irrelevant factor $1/(2 \pi \alpha')$)
\begin{align}
 \delta S = \int d^2 \sigma \delta L
   =- \frac12 \int_{\del \Sigma} d\tau &\left[
 \eta_{+}\left(g_{IJ}\psi_{-}^{I}\del_{+} x^{J}
         + \psi_{+}^{I} \del_{I} U(x)\right) \right. \nn
 &\qquad \left. + \eta_{-}\left( - g_{IJ}\psi_{+}^{I}\del_{-} x^{J}
          + \psi_{-}^{I} \del_{I} U(x)\right)
\right]~.
\end{align}
In order to preserve the ${\cal N}=1$ supersymmetry, 
say $\eta_{+} = \eta_{-}$, 
we have to assign the boundary condition to the fields.
By assigning $\eta_{+} = \eta_{-}$, we find
\begin{align}
 \delta S = -\frac12 \int_{\del \Sigma} d\tau \;\eta_{+} &\left[
g_{IJ}(\psi_{-}^{I}-\psi_{+}^{I})\del_0  x^{J} \right. \nn
&  \left. +g_{IJ}(\psi_{-}^{I}+\psi_{+}^{I})\del_1 x^{J}
 + (\psi_{-}^{I}+\psi_{+}^{I})\del_I U(x)
\right]~.
\end{align}
Therefore the boundary conditions are
$\psi^I = \bar{\psi}^I$ for the tangent direction
of the brane and $\psi^I = - \bar{\psi}^I$ for the normal direction.
Furthermore, $U$ must be constant along the tangent direction 
$\partial_I U = 0$.

Next we compare this result with the one from Killing spinor analysis.
Here we should note that the Killing spinors of $(1,1)$ case 
(\ref{11killing}) are given by replacing the ones of $(2,2)$
case (\ref{22killing}) of $\varphi_{j\bar{k}}=0$ with 
$\alpha \to - \zeta$, $ i \partial_{j} W \to \partial_{j} U $ and
$-i \partial_{\jb} \bar{W} \to \partial_{\jb} U $.
For the D-branes wrapped on complex submanifolds,
we can construct supersymmetric D-branes when
$p=1,5,9$ since the condition $\alpha = - \zeta$ is
compatible only in these cases.
By using (\ref{holomorphic}), we obtain 
$\zeta = a - ia, a + ia, a - ia$ with real $a$ for $p =
1,5,9$, respectively and
\begin{equation}
 \partial_I U = 0 ~,~~ I : \mbox{tangent} ~.
\label{11spc}
\end{equation}
This condition is the same as the one obtained from the string worldsheet
 analysis.
The D-branes wrapped on special Lagrangian submanifolds are
 also examined and they are supersymmetric if
$\zeta =a +ia$ for a real parameter $a$
and superpotential $U$ satisfy (\ref{11spc}).
Therefore we conclude that the Killing spinor analysis reproduces the
result from the string worldsheet analysis.

\section{Conclusion and discussions}
\label{section-conclusion}

In this paper, we have investigated the D-branes in the supersymmetric
pp-wave backgrounds constructed in \cite{Maldacena:2002fy}.
The corresponding open string worldsheet theories in the
lightcone gauge are supposed to be the Landau-Ginzburg theories on the
two dimensional worldsheet with boundary. 
In the D-brane worldvolume analysis, the supersymmetry can be examined
by using the kappa symmetry projection.
The results are compared to the D-branes in the Landau-Ginzburg models.

For the ${\cal N}=(2,2)$ supersymmetric case without holomorphic Killing
vector terms and gauge field excitations, 
we obtain the two types of supersymmetric
D-branes. One is called as A-type D-brane, which is wrapped on a
special Lagrangian submanifold and the other is called as B-type
D-brane, which is wrapped on a complex submanifold.
Moreover, there are conditions on the superpotential in both cases.
We have shown that these results can be reproduced by
the analysis of Landau-Ginzburg models \cite{Hori:2000ck}.
As for an A-type D-brane, we obtain the BPS D-branes with the non-constant
imaginary part of the superpotential by including non-trivial gauge fields.
The D-branes on the backgrounds with holomorphic Killing vector terms
and the ${\cal N}=(1,1)$ backgrounds are also studied.

The correspondence between the spacetime Killing spinor
analysis and the Landau-Ginzburg model analysis seems work quite well,
nevertheless, there is a disagreement.
In the D-brane worldvolume analysis, the A-type D-branes should
be wrapped on a {\it special } Lagrangian submanifold, however in the
string worldsheet analysis, the A-type D-branes should be 
wrapped on a Lagrangian submanifold.
This discrepancy may originate from the fact that the Killing spinor
analysis use the {\it spacetime} spinors, however, the Landau-Ginzburg models
have the {\it worldsheet} spinors.
In the ${\cal N}=(2,2)$ superconformal field theory, there is a spectral flow
symmetry and it is believed that it relates 
the spacetime supersymmetry to the worldsheet one.
In the superconformal case like \cite{Ooguri:1996ck,Becker:1996ay},
we have to assign the boundary condition also for the spectral flow operator
in order to reproduce the spacetime analysis.
In our lightcone analysis, there must be much closer relation between
the spacetime and worldsheet supersymmetry, thus we cannot use the same
analysis.
However, it is natural to expect that we resolve this problem 
if we assign an alternative condition corresponding to
the requirement of the {\it superconformal} symmetry, which we have not
known yet.
It is important to investigate this aspect more closely\footnote{
In the $U(4)$ formalism \cite{Berkovits:2002vn} of the covariant
gauge, the D-branes correspond to the boundary conditions which preserve
superconformal symmetry. 
This construction of the D-branes may resolve this problem.
We would like to thank Nathan Berkovits for the useful comment.}.

Although we have used the flat transverse space throughout this paper,
we can also treat the curved (Calabi-Yau) transverse space in the same manner.
We can take the local frame of the Calabi-Yau space as follows:
\begin{align}
 &ds^2_{8}=2 g_{i\jb}e^{i}e^{\jb}~,\quad \omega=i g_{i\jb}e^{i}\wedge e^{\jb}\ 
 (\text{K\"ahler form})~,\nn
 &\Omega=4 e^{1}\wedge e^{2}\wedge e^{3}\wedge e^{4} \ 
   (\text{Holomorphic 4-form})~,\nn
 & \nabla \epsilon =d\epsilon + \boldsymbol{\omega}^{i\jb}\Gamma_{i\jb}\epsilon
  \ (\text{Covariant derivative})~,\quad 
 \boldsymbol{\omega}^{i}{}_{i}=0~,
\end{align}
where $g_{i\jb}$ is the {\em flat} K\"ahler metric defined
as $(g_{i\jb})={\rm diag(1,1,1,1)}$,  $e^{i}, e^{\jb}$ are the vielbeins and
$\boldsymbol{\omega}^{i\jb}$
is the spin connection. In this frame, two covariantly constant
spinors in this Calabi-Yau 4-fold can be expressed as $\vac$ and $\vact$
which satisfy $d \vac= d \vact=0$ and the features in Appendix A.
We can consider the pp-wave background with this Calabi-Yau 
4-fold \cite{Maldacena:2002fy} in this frame.
The Killing spinor can be written just
the same form as eq.(\ref{22killing}). All the analyses of section 
\ref{section-D-branes} and \ref{section-general-backgrounds} can be repeated
just the same way by replacing $dz^{i}\to e^{i},\ d\bar z^{\ib}\to e^{\ib}$.
Therefore, the same result is obtained also in the case of curved transverse
space.

In order to use the transverse space as $\mathit{Spin}(7)$ or $G_2$ manifold, 
we have to use the real coordinates instead of the complex coordinates. 
In this case, we may find the supersymmetric D-brane wrapped on a Cayley
cycle as in \cite{Becker:1996ay} and hence it is worthwhile to study it.

It is also important to extend to the more general setups.
The pp-wave with constant 3-forms corresponds to the Penrose limit
of $AdS_3 \times S^3 \times T^4 \ (\mbox{or } K3)$ 
\cite{Hikida:2002in,Lunin:2002fw,Gomis:2002qi,Hikida:2002xu,Gava:2002xb}
and it was shown that, in the case with
non-constant 3 forms, there is no supersymmetry
linearly realized on the worldsheet of the lightcone
gauge \cite{Russo:2002qj}. 
It is interesting to consider the D-branes in this background because
the condition of consistent D-branes is supposed to be different from ours.

In the Landau-Ginzburg description, we can apply the mirror symmetry to
the D-branes \cite{Hori:2000kt,Hori:2000ck}. 
It is important to see how the mirror
symmetry act on the D-brane configurations of 
the string theories in the covariant gauge.



\subsection*{Acknowledgements}
We would like to thank Tohru Eguchi, Isao Kishimoto, Takeo Inami,
Kazuhiro Sakai, Yuji Sugawara
and Hiromitsu Takayanagi for useful discussions and comments.
S.~Y. is supported in part by Soryushi Shogakukai.


\appendix
\section{Gamma matrices and useful formulae}

We use the convention of Gamma matrices as
\begin{equation}
 \{\Gamma^{+},\Gamma^{-}\}=-2~,\quad \{\Gamma_{+}, \Gamma_{-}\}=-2~, \quad
 \{\Gamma^{i},\Gamma^{\ib}\}=2g^{i\ib}~,\quad    
           \{\Gamma_{i},\Gamma_{\ib}\}=2g_{i\ib} ~.
\end{equation}
We also define
\begin{equation}
\Gamma_{+-}:=\frac12(\Gamma_+ \Gamma_- - \Gamma_- \Gamma_+)~,\quad
  \Gamma_{i\ib}:=\frac12(\Gamma_i \Gamma_{\ib} - \Gamma_{\ib} \Gamma_i)~,
\end{equation}
and $\Gamma_{m_1 \cdots m_n}$ in the similar way.
Using this notation, we can show
\begin{eqnarray}
 &&(\Gamma_{+-})^2=1~,\quad
   \Gamma_{+-}\Gamma_{-}=-\Gamma_{-}\Gamma_{+-}~,\quad
  \Gamma_{+-}\Gamma_{+}=-\Gamma_{+}\Gamma_{+-}~,\nn
 &&[\Gamma_{i\ib},\Gamma^{\kb}]=2\Gamma_{i}\delta^{\kb}_{\ib}~,\quad
  [\Gamma_{i\ib},\Gamma^{k}]=-2\Gamma_{\ib}\delta^{k}_{i}~.\nn
 &&A^{i}\bar A^{\ib}A^{j}\bar A^{\jb}\Gamma_{i\ib}\Gamma_{j\jb}
   =|A|^4 ~, \quad |A|^2 := A^{i}\bar A^{\ib}g_{i\ib}~,
\end{eqnarray}
where $A^i$ are arbitrary vectors.
In order to express the spinors it is convenient to use the fock space
formalism. The vacua are given by
\begin{eqnarray}
 &&\Gamma_{+}\vac=\Gamma^{m} \vac=\Gamma_{\mb} \vac=0~,\nn
 &&\Gamma_{+}\vact=\Gamma^{\mb} \vact=\Gamma_{m} \vact=0~,
\qquad \vact
 =\frac14 \Gamma^{\bar 1}\Gamma^{\bar 2}\Gamma^{\bar 3}\Gamma^{\bar 4}\vac~,
\end{eqnarray}
and satisfy
\begin{equation}
 \Gamma_{i\ib}\vac=-g_{i\ib}\vac~,\quad
        \Gamma_{i\ib}\vact=+g_{i\ib}\vact~,\quad
 \Gamma_{+-}\vac=-\vac~,\quad \Gamma_{+-}\vact=-\vact~.
\end{equation}


\section{Supersymmetric Landau-Ginzburg models}
\subsection{${\cal N}= (2,2)$ case}
\label{22LG}

Let $(\sigma^{0}=\tau, \sigma^{1}=\sigma)$ be
  the coordinates of the two dimensional Minkowski space.
It is convenient to use
\begin{align}
 \sigma^{\pm} = \frac{1}{2} (\tau \pm \sigma) ~,~~
 \del_{\pm}:=\deldel{}{\sigma^{\pm}}=\del_{\tau}\pm \del_{\sigma}~.
\end{align}
We introduce the fermionic coordinates $(\ta^{\pm},\tb^{\pm})$, and
define the supertranslation generator and supercovariant derivative as
\begin{align}
 &Q_{\pm}=\deldel{}{\ta^{\pm}} + i \tb^{\pm}\del_{\pm}~,\qquad
 \Qb_{\pm}=-\deldel{}{\tb^{\pm}} - i \ta^{\pm}\del_{\pm}~,\nn
 &D_{\pm}=\deldel{}{\ta^{\pm}} - i \tb^{\pm}\del_{\pm}~,\qquad
 \Db_{\pm}=-\deldel{}{\tb^{\pm}} + i \ta^{\pm}\del_{\pm}~.
\end{align}
Anti-commutators between these differential operators become
\begin{equation}
 \{Q_{\pm},\Qb_{\pm}\}=-2i \del_{\pm}~,\quad
 \{D_{\pm},\Db_{\pm}\}= 2i \del_{\pm}~,\quad
  (\text{others})=0~.
\end{equation}

We use chiral superfield $\Phi^{i}$ and its complex 
conjugate $\Phb^{\ib}$ satisfying
\begin{align}
 \Db_{\pm} \Phi^{i}=0~,\qquad  D_{\pm} \Phb^{\ib}=0~.
\end{align}
These chiral superfields are expanded by fermionic coordinates as
\begin{align}
 \Phi^{i} &=\pha^{i}+\sqrt2 \ta^{+}\psi_{+}^{i}+\sqrt2 
                  \ta^{-}\psi_{-}^{i}+2\ta^{+}\ta^{-}F^{i}
      -i\ta^{+}\tb^{+}\del_{+}\pha^{i}-i\ta^{-}\tb^{-}\del_{-}\pha^{i}\nn
  &\quad   -i\sqrt2 \ta^{+}\ta^{-}\tb^{-}\del_{-}\psi_{+}^{i}
     -i\sqrt2 \ta^{-}\ta^{+}\tb^{+}\del_{+}\psi_{-}^{i}
     -\ta^{+}\ta^{-}\tb^{-}\tb^{+}\del_{+}\del_{-}\pha^{i}~,\nn
 \Phb^{\ib}&=\phb^{\ib}-\sqrt2 \tb^{+}\psb_{+}^{\ib}-\sqrt2 
                  \tb^{-}\psb_{-}^{\ib}+2\tb^{-}\tb^{+}\Fb^{\ib}
      +i\ta^{+}\tb^{+}\del_{+}\phb^{\ib}
      +i\ta^{-}\tb^{-}\del_{-}\phb^{\ib}\nn
  &\quad   -i\sqrt2 \ta^{-}\tb^{-}\tb^{+}\del_{-}\psb_{+}^{\ib}
     -i\sqrt2 \ta^{+}\tb^{+}\tb^{-}\del_{+}\psb_{-}^{\ib}
     -\ta^{+}\ta^{-}\tb^{-}\tb^{+}\del_{+}\del_{-}\phb^{\ib}~.
\end{align}
The action we consider is the ${\cal N} =(2,2)$  Landau-Ginzburg
models (\ref{LG}).  Integrating fermionic
coordinates, the K\"ahler potential term becomes\footnote{
We concentrate on only the flat target space.}
\begin{align}
 L_K&=\int d^4\ta g_{i\jb}\Phi^{i}\Phb^{\jb} :=
     \frac14 \int d\ta^{+} d\ta^{-} d \tb^{-}
                  d\tb^{+}g_{i\jb}\Phi^{i}\Phb^{\jb}\nn
 &=\frac12 g_{i\jb} (\del_+ \pha^{i}\del_- \phb^{\jb} 
                   + \del_+ \phb^{\jb}\del_- \pha^{i}
  +i\psb_{+}^{\jb} \dlr_{-}\psi_{+}^{i} +i\psb_{-}^{\jb} \dlr_{+}\psi_{-}^{i})
  +g_{i\jb}F^{i}\Fb^{\jb}~.
\end{align}
The superpotential term can be calculated as 
\begin{align}
 L_W&= \frac12 \left(\int d^2\ta W(\Phi)+(c.c.)\right)
 :=\frac14 \int d\ta^{-}d\ta^{+}W(\Phi)|_{\tb^{\pm}=0}+(c.c.)\nn
 &=\frac12\del_{i} W(\pha) F^{i} +\frac12\del_{\ib} \Wb(\phb)\Fb^{\ib} 
-\frac12\del_{i}\del_{j}W(\pha)\psi_{+}^{i}\psi_{-}^{j}
-\frac12\del_{\ib}\del_{\jb}\Wb(\phb)\psb_{-}^{\ib}\psb_{+}^{\jb}~.
\end{align}
The holomorphic Killing vector term is given by 
\cite{Bagger:1982fn,Alvarez-Gaume:1983ab,Gates:1984py}
\begin{align}
 L_{V}=-|m|^2 g_{i\jb}V^{i}V^{\jb}
 -\frac{i}{2}(g_{i\ib}\del_{j}V^{i}-g_{j\jb}\del_{\ib}V^{\jb})
  (m\psb_{-}^{\ib}\psi_{+}^{j}+\mb\psb_{+}^{\ib}\psi_{-}^{j}) ~,
\end{align}
where the holomorphic Killing vector $V^{i}$ satisfies
\begin{align}
 &V_{i}=i\vp_{i\jb}\zb^{\jb}~,\quad V_{\ib}=-i\vp^*_{\ib j}z^{j}~,
     \quad V^{i}\del_{i}W =0~,\nn
 &\del_{i}V^{\jb}=0~, \quad \del_{\jb}V^{i}=0~,\quad 
 \del_{i}V_{\jb}=-\del_{\jb}V_{i}=(\text{constant})~.
\end{align}
By using the equation of motion, we can set $F^i$ as 
\begin{equation}
 F^i=-\frac12g^{i\ib}\del_{\ib}\Wb(\phb)~,
\end{equation}
and we obtain the total Lagrangian
\begin{eqnarray}
\lefteqn{L=L_K+L_W+L_V}\nn
&&=\frac12 g_{i\jb}
   (\del_+ \pha^{i}\del_- \phb^{\jb} + \del_+ \phb^{\jb}\del_- \pha^{i} 
  +i\psb_{+}^{\jb} \dlr_{-}\psi_{+}^{i}
  +i\psb_{-}^{\jb} \dlr_{+}\psi_{-}^{i})\nn
&& -\frac12\del_{i}\del_{j}W(\pha)\psi_{+}^{i}\psi_{-}^{j}
        -\frac12\del_{\ib}\del_{\jb}\Wb(\phb)\psb_{-}^{\ib}\psb_{+}^{\jb}
        -\frac14g^{i\jb}\del_{i} W(\pha)\del_{\jb}\Wb(\phb)\nn
&&-|m|^2 g_{i\jb}V^{i}V^{\jb}
 -\frac{i}{2}(g_{i\ib}\del_{j}V^{i}-g_{j\jb}\del_{\ib}V^{\jb})
  (m\psb_{-}^{\ib}\psi_{+}^{j}+\mb\psb_{+}^{\ib}\psi_{-}^{j}) ~.
\end{eqnarray}

The supersymmetry transformation on this action 
can be described by using
the two complex fermionic parameter $\eta_{+}, \eta_{-}$ as
\begin{align}
 &\delta z^{i}=\eta_{+}\psi_{-}^{i}-\eta_{-}\psi_{+}^{i}~,
 &&\delta \zb^{\ib}=-\etb_{+}\psb_{-}^{\ib}+\etb_{-}\psb_{+}^{\ib}~,\nn
 &\delta \psi_{+}^{i}=i\etb_{-}\del_{+}z^{i}
                 -\frac12 \eta_{+}g^{i\jb}\del_{\jb}\Wb
         -i\etb_{+}\mb V^{i}~,
 &&\delta \psb_{+}^{\ib}=-i\eta_{-}\del_{+}\zb^{\ib}
                 -\frac12 \etb_{+}g^{\ib j}\del_{j}W
         +i\eta_{+}m V^{\ib}~,
\nn
 &\delta \psi_{-}^{i}=-i\etb_{+}\del_{-}z^{i}
                 -\frac12 \eta_{-}g^{i\jb}\del_{\jb}\Wb
         +i\etb_{-} m V^{i}~,
 &&\delta \psb_{-}^{\ib}=i\eta_{+}\del_{-}\zb^{\ib}
                 -\frac12 \etb_{-}g^{\ib j}\del_{j}W
         -i\eta_{-} \mb V^{\ib}~.
\end{align}
The variation of the Lagrangian becomes
\begin{align}
 \delta L&=\frac12 g_{i\jb}\Big[
  \eta_{+}\del_{-}(\del_{+}\phb^{\jb}\psi^{i}_{-})
  -\eta_{-}\del_{+}(\del_{-}\phb^{\jb}\psi^{i}_{+})
  -\etb_{+}\del_{-}(\del_{+}\pha^{i}\psb^{\jb}_{-})
  +\etb_{-}\del_{+}(\del_{-}\pha^{i}\psb^{\jb}_{+})
  \Big]\nn
 &\quad -\frac{i}{4}\Big[
      \eta_{+}\del_{-}(\del_{\ib}\Wb\psb^{\ib}_{+})
     +\eta_{-}\del_{+}(\del_{\ib}\Wb\psb^{\ib}_{-})
     +\etb_{+}\del_{-}(\del_{ i } W \psi^{ i }_{+})
     +\etb_{-}\del_{+}(\del_{ i } W \psi^{ i }_{-})
    \Big]\nn
 &\quad -\frac12\Big[
  \eta_{+} m \del_{-}(V_{ i }\psi^{ i }_{+})
 -\eta_{-}\mb\del_{+}(V_{ i }\psi^{ i }_{-})
 -\etb_{+}\mb\del_{-}(V_{\jb}\psb^{\jb}_{+})
 +\etb_{-} m \del_{+}(V_{\jb}\psb^{\jb}_{-})
   \Big] ~.
\label{22variation}
\end{align}
This is a total derivative form 
and hence there is $\Ncal=(2,2)$ supersymmetry if there is no boundary.

\subsection{${\cal N}= (1,1)$ case}
\label{11LG}

Let us introduce the two pure imaginary 
fermionic coordinates $(\ta^{+},\ta^{-})$,
then the supertranslation generator
and supercovariant derivative can be defined as
\begin{equation}
 Q_{\pm}=\deldel{}{\ta^{\pm}}-i\ta^{\pm}\del_{\pm}~,\qquad
 D_{\pm}=\deldel{}{\ta^{\pm}}+i\ta^{\pm}\del_{\pm}~.
\end{equation}
A real superfield $\Phi^I$ can be expanded as
\begin{align}
 &\Phi^{I}=x^{I}+\ta^{+}\psi_{+}^{I}
          +\ta^{-}\psi_{-}^{I}+\ta^{+}\ta^{-}F^{I}~.
\end{align}
The Lagrangian of ${\cal N}=(1,1)$ Landau-Ginzburg models 
can be written as (\ref{N=1action})
\begin{align}
 L &=\int d^2 \ta \left[\frac12 g_{IJ} D_{+}\Phi^{I} D_{-}\Phi^{J}
                          + i U(\Phi)\right]\nn
   &=\frac12 g_{IJ}\left[\del_{+}x^{I} \del_{-}x^{J}
           +i \psi_{+}^{I}\del_{-}\psi_{+}^{J}
           +i \psi_{-}^{I}\del_{+}\psi_{-}^{J}
        -F^{I}F^{J} \right] \nn
        & \qquad  - i \del_{I} U(x) F^{I}
      +i  \del_{I}\del_{J} U(x) \psi_{+}^{I} \psi_{-}^{J}  ~.
\end{align}
Eliminating $F^I$ by
\begin{equation}
 F^{I}=- i \del^{I} U(x)~,
\end{equation}
we obtain
\begin{align}
 L&=\frac12 g_{IJ}\left[\del_{+}x^{I} \del_{-}x^{J}
           +i \psi_{+}^{I}\del_{-}\psi_{+}^{J}
           +i \psi_{-}^{I}\del_{+}\psi_{-}^{J} \right] \nn
      & \qquad  + i \del_{I}\del_{J} U(x) \psi_{+}^{I} \psi_{-}^{J}
      - \frac12 \del_{I}U(x) \del^{I}U(x) ~.
\end{align}

The supersymmetry transformation is expressed with two real parameters
$\eta_{\pm}$ as 
\begin{align}
 &\delta x^{I}=\eta_{+} \psi_{-}^{I} + \eta_{-} \psi_{+}^{I}~,\nn
 &\delta \psi_{+}^{I} = i\eta_{-}\del_{+} x^{I}
                               - i \eta_{+} \del^{I} U(x)~,\nn
 &\delta \psi_{-}^{I} = i\eta_{+}\del_{-} x^{I}
                               + i \eta_{-} \del^{I} U(x)~.
\end{align}
The variation of Lagrangian can be written as
\begin{align}
 \delta L =
  \frac12 \eta_{+} \del_{-}\left(g_{IJ}\psi_{-}^{I}\del_{+}x^{J}
         + \psi_{+}^{I} \del_{I} U(x)\right)
 +\frac12 \eta_{-} \del_{+}\left(g_{IJ}\psi_{+}^{I}\del_{-}x^{J}
         - \psi_{-}^{I} \del_{I} U(x)\right).
\label{11variation}
\end{align}
This is the total derivative form and hence there is $\Ncal=(1,1)$
supersymmetry if there is no boundary on worldsheet.



\providecommand{\href}[2]{#2}\begingroup\raggedright\endgroup
\end{document}